\documentclass{ws-ijmpd}
\usepackage[]{amssymb}
\usepackage[]{rotating}
\begin{document}

\markboth{B. Arbutina} {Constraints on the Massive Supernova
Progenitors}

\catchline{}{}{}{}{}
%
%%%%%%%%%%%%%%%%%%%%%%%%%%%%%%%%%%%%%%%%%%%%%%%%%%%%%%%%%%%%%%%%%%%%

\title{CONSTRAINTS ON THE MASSIVE SUPERNOVA
PROGENITORS  }

\author{BOJAN ARBUTINA$^{1, 2}$}

\address{$^1$Astronomical Observatory, Volgina 7, 11160 Belgrade, Serbia\\
$^2$Department of Astronomy, Faculty of Mathematics, University of Belgrade, Studentski trg 16, 11000 Belgrade, Serbia\\
barbutina@aob.bg.ac.yu}

\maketitle

\begin{history}
\received{Day Month Year}
\revised{Day Month Year}
\comby{Managing Editor}
\end{history}

\begin{abstract}
Generally accepted scheme distinguishes two main classes of
supernovae (SNe): Ia resulting from the old stellar population
(deflagration of a white dwarf in close binary systems), and SNe
of type II and Ib/c whose ancestors are young massive stars (died
in a core-collapse explosion). Concerning the latter, there are
suggestions that the SNe II are connected to early B stars, and
SNe Ib/c to isolated O or Wolf-Rayet (W-R) stars. However, little
or no effort was made to further separate SNe Ib from Ic. We have
used assumed SN rates for different SN types in spiral galaxies in
an attempt to perform this task. If isolated progenitor hypothesis
is correct, our analysis indicates that SNe Ib result from stars
of main-sequence mass  $23 \mathcal{M}_{\odot} \lesssim
 \mathcal{M} \lesssim 30 \mathcal{M}_{\odot}$, while
the progenitors of SNe Ic are more massive stars with $\mathcal{M}
\gtrsim 30 \mathcal{M}_{\odot}$.
 Alternatively, if the majority of SNe Ib/c appear in close binary systems (CBs)
 then they would result from the same progenitor population as
 most of the
 SNe II, i.e. early B stars with initial masses of order
 $\mathcal{M} \sim 10 \mathcal{M}_{\odot}$.
Future observations of SNe at high-redshift ($z$) and their rate
will provide us with unique information on SN progenitors and
star-formation history of galaxies. At higher-$z$ (deeper in the
cosmic past) we expect to see the lack of type Ia events, i.e. the
dominance of core-collapse SNe. Better understanding of the
stripped-envelope SNe (Ib/c), and their potential use as distance
indicators at high-$z$, would therefore be of great practical
importance.
\end{abstract}


\section{Introduction}

Interest in the study of supernovae (SNe) has been significantly
increased following the renewed searches for SNe in recent years.
Now more than 3000 SNe have been discovered since 1885, many of up
to them at cosmological distances. Much effort has been invested
in the classification of SNe, in understanding their progenitors
and determining SN rate (and rates for different SN and parent
galaxy types). Consequently we arrived at the generally accepted
scheme with two distinctive classes of SNe: Ia resulting from the
old stellar population (deflagration of a C-O white dwarf in close
binary systems), and SNe of type II and Ib/c whose ancestors are
young massive stars (died in a core-collapse explosion).

Historically, classification of SNe, according to their optical
spectra, began by recognizing SNe I, with no hydrogen lines, and
SNe II which do show hydrogen in their spectra. In addition, SNe
II were shown to exhibit much wider photometric behavior than SNe
I, which seemed to be a rather homogeneous class of objects.
Nevertheless, it was shown later that there are actually two
spectroscopically and photometrically distinct subclasses of SNe
I: Ia which were only located in ellipticals, and Ib found in HII
regions and spiral arms, which strongly suggested that their
progenitors were massive young stars with their hydrogen envelopes
stripped. The third subclass, SNe Ic, discovered later, show no
helium lines either, and thus corresponds to the massive stars
stripped of their H and He envelopes. For more detailed review see
Refs. 1--4.

There are suggestions that the SNe II are connected to early B
stars, and SNe Ib/c to isolated O or Wolf-Rayet (W-R) stars.
However, little or no effort was made to further separate SNe Ib
from Ic. In the subsequent analysis we will use assumed SN rates
for different SN types in spiral galaxies in an attempt to perform
this task. An alternative, according to which all core-collapse
SNe result from the same progenitor population is also discussed.

\section{Analysis}

\subsection{Masses of supernova progenitors}

The first constraints on the masses of supernova progenitors were
put by Kennicutt\cite{5}, and later by van den Bergh\cite{6}, from
the assumed supernova and star formation rates (SNR and SFR,
respectively) for the core-collapse supernovae and their
progenitors. Because of the short life of massive stars, the
number of core-collapse events per century should be equal to the
number of new born stars in the same time period, within the
appropriate mass range,
\begin{equation}
\mathrm{SNR} = \mathrm{SFR,}
\end{equation}
when converted to appropriate units. Star formation rate for the
stars in the mass range from $\mathcal{M}_L$ to $\mathcal{M}_U$ is
given with
\begin{equation}
\mathrm{SFR} \propto \int_{\mathcal{M}_L}^{\mathcal{M}_U}
f(\mathcal{M})d\mathcal{M},
\end{equation}
where $f$ is the initial mass function (IMF), $f(\mathcal{M})=A
\mathcal{M} ^{-\beta}$ in Salpeter's form\cite{7}.

According to Ref. 6 progenitors of SN II have masses $8
\mathcal{M}_{\odot} \lesssim
 \mathcal{M}_{\mathrm{II}} \lesssim 18 \mathcal{M}_{\odot}$, corresponding to early B stars,
  whereas isolated O or W-R stars with $\mathcal{M}_{\mathrm{Ib}} \gtrsim 18
 \mathcal{M}_{\odot}$ become SN Ib. SN Ic were not recognized
 separately from Ib at that time.\footnote{In his study van den Bergh
 used stripped-envelope (Ib) to total core-collapse (Ib+II) ratio $\nu_{\mathrm{se}} /\nu _{\mathrm{cc}} \approx
 0.26$.\cite{6}}

 In order to calculate the limiting masses for SN Ib/c progenitors
 from the mass spectrum of star formation we need to know the IMF
 and the SNRs for these particular classes of SNe. We will adopt, as the authors cited, the Miller--Scalo mass function with $\beta =2.5$,\cite{8}
 upper mass limit for the core-collapse SN progenitors of
 $\mathcal{M}_U=100 \mathcal{M}_{\odot}$, and the fixed predetermined
 lower limit of $\mathcal{M}_L=8 \mathcal{M}_{\odot}$,
 consistent with theory.\cite{1,9} Recently determined
 supernova rates, i.e relative numbers that we are interested in,
 are given in Table 1.

 SNRs for a given SN and galaxy type is defined as:
\begin{equation}
\nu=\frac{N}{T}\ \ \mathrm{[ SNu ]} ,
\end{equation}
where $N$ is the number of SNe discovered in a given sample of
galaxies during the total control time $T$, and supernova unit is
$1\ \mathrm{SNu} = \mathrm{SNe}$ per $10^{10} L_{\odot}$ per
century. Total control time incorporates galaxy luminosity as a
normalization factor, since it has been shown that it correlates
with the SNR, and the probability of SN detection, depending on
the photometric properties of SN type in question (see Refs. 3,
10).

Our intent was to use entirely the study given in Ref. 10
 since it provides a well defined control
time. Control time is of utmost importance for correct
determination of the SNR, which implies exclusion of all SNe
discovered by chance for which $T$ is undefinable. However,
statistics were not large enough to separate stripped-envelope
SNe, that, for this reason, were lumped together as Ib/c in Refs.
10 and 11.

Nonetheless, the majority of SNe Ib/c, as new types, were
discovered in systematic SN searches, which means that, although
unknown to us, there is a control time calculable. The crucial
thing is that this control time is the same for SNe Ib and Ic,
which are photometrically indistinguishable (or, at least, very
much alike)! In other words, the selection effects acting on SNe
Ib/c would be the same and therefore, for this initial study, we
can assume
\begin{equation}
\frac{\nu_{\mathrm{Ib}}}{\nu_{\mathrm{Ic}}}=
\frac{N_{\mathrm{Ib}}}{N_{\mathrm{Ic}}}\approx
\frac{\bar{N}_{\mathrm{Ib}}}{\bar{N}_{\mathrm{Ic}}},
\end{equation}
where $\bar{N}$ is the total number of recorded SNe. As a database
we used the October 2004 version of the Asiago Supernova Catalogue
(hereafter ASC).\cite{12} A cutoff at redshift $z=0.03$ has been
induced to make the results consistent with the rates in the local
universe. The resulting numbers are given in the last column of
Table 1.

If we adopt that approximately 83 per cent of all core-collapse
events are SNe II and only 6 and 11 per cent are SNe Ib and Ic
respectively, from
\begin{equation}
\frac{\int_{\mathcal{M}_L}^{\bar{\mathcal{{M}}}_{\mathrm{Ib}}}
\mathcal{M} ^{-\beta}d\mathcal{M}
}{\int_{\mathcal{M}_L}^{\mathcal{M}_U} \mathcal{M}
^{-\beta}d\mathcal{M}}=\frac{\nu_{\mathrm{II}}}{\nu_{\mathrm{cc}}}
\end{equation}
\begin{equation}
\frac{\int_{\bar{\mathcal{{M}}}_{\mathrm{Ic}}}^{\mathcal{M}_U}
\mathcal{M} ^{-\beta}d\mathcal{M}
}{\int_{\mathcal{M}_L}^{\mathcal{M}_U} \mathcal{M}
^{-\beta}d\mathcal{M}}=\frac{\nu_{\mathrm{Ic}}}{\nu_{\mathrm{cc}}}
\end{equation}
we obtain $\bar{\mathcal{M}}_{\mathrm{Ib}} \approx 24
\mathcal{M}_{\odot}$ and $\bar{\mathcal{M}}_{\mathrm{Ic}} \approx
31 \mathcal{M}_{\odot}$.

 \begin{table}[ph]
\tbl{Supernova rates in SNu for different galaxies and SN types
from Ref. 10. $h$ is the Hubble parameter $h=H_o/(75\ \mathrm{km\
s^{-1} Mpc^{-1}})$.
  Last but one column is the stripped-envelope (Ib/c) relative to total core-collapse SN rate
  (Ib/c+II). The last column shows the SN Ib to Ic ratio found in this study. }
 {\begin{tabular}{@{\extracolsep{6.0mm}}lcccc@{}}
  \hline
   Galaxy & \multicolumn{2}{c}{$\nu \ \mathrm{[SNu]} $ }&$\nu_{\mathrm{se}} /\nu _{\mathrm{cc}}$& $\bar{N}_{\mathrm{Ib}}/\bar{N}_{\mathrm{Ic}}$ \\
   type            &  Ib/c & II & & \\
   \hline
   E--S0     &  $<0.01 h^2$         & $<0.02 h^2$      &--  & --   \\
   S0a--Sb   &  $0.11\pm0.06\ h^2$  &$0.42\pm0.19\ h^2$&0.21& 0.92$^{c}$ \\
   Sbc--Sd   &  $0.14\pm0.07\ h^2$  &$0.86\pm0.35\ h^2$&0.14& 0.45\ \ \\
   Others$^{a}$&$0.22\pm0.16\ h^2$  &$0.65\pm0.39\ h^2$&0.25& 0.42\ \ \\
   All       &  $0.08\pm0.04\ h^2$  &$0.40\pm0.19\ h^2$&0.17& 0.53$^{b}$ \\
 \hline
 \vspace{0.1mm}
\end{tabular}}
\begin{minipage}{0.85\textwidth}
\begin{flushleft}
  $^{a}$Others include types Sm, Irregulars and Peculiars.\\
  $^{b}$The ratio for $i\leq 45^{\circ}$ is 0.51, and 0.55 for $i> 45^{\circ}$, and thus it is not
   effected by galaxy inclination.\\
  $^{c}$This high value is probably a consequence of the small-number
  statistics.
\end{flushleft}
\end{minipage}
\end{table}

If one uses near infrared (K band) instead of B luminosity of a
galaxy as a better tracer of its stellar mass, then supernova rate per unit mass (SNuM) can be calculated.\cite{13}
Initial mass values for the progenitors of different types of
core-collapse SNe as a function of the host galaxy's Hubble type,
obtained by using rates in SNuM from Ref. 13, are given in Table 2.
The numbers in Table 2 are probably too limited by a small number statistics to be useful, but the mean values $\bar{\mathcal{M}}_{\mathrm{Ib}} \approx 22
\mathcal{M}_{\odot}$, $\bar{\mathcal{M}}_{\mathrm{Ic}} \approx
29 \mathcal{M}_{\odot}$, roughly match those obtained previously.

 \begin{table}[ph]
\tbl{Initial mass values for the progenitors of different types of
core-collapse SNe as a function of the host galaxy's Hubble type,
obtained by using rates from Ref. 13.}
 {\begin{tabular}{@{\extracolsep{-2.0mm}}lcccccccc@{}}
  \hline
   Galaxy  &  $\nu_{\mathrm{se}} /\nu _{\mathrm{cc}}$ && ${\mathcal{M}}_{\mathrm{II}}$ [$\mathcal{M}_{\odot}$]
          && ${\mathcal{M}}_{\mathrm{Ib}}$ [$\mathcal{M}_{\odot}$] && ${\mathcal{M}}_{\mathrm{Ic}}$ [$\mathcal{M}_{\odot}$] &  \\
   type      &      && $\overbrace{\hspace{15mm}}$ && $\overbrace{\hspace{15mm}}$ &&  $\overbrace{\hspace{15mm}}$ &  \\
   \hline
   E--S0     & 0.23 & 8 &  & 20 &  & 30 &  & 100  \\
   S0a--Sb   & 0.14 & 8 &  & 27 &  & 34 &  & 100  \\
   Sbc--Sd   & 0.24 & 8 &  & 20 &  & 24 &  & 100  \\
  \hline
 \vspace{0.1mm}
\end{tabular}}
\end{table}

These initial mass values correspond to the massive short-lived O
stars. If these are truly progenitors of SNe Ib/c they will die
not far from the place where they were born. Table 3 gives the
number of O stars of different subclass out of, or in the H II
regions. It shows that early O stars are more likely to be located
in H II regions than the late O stars, with O8 being the
intermediate class. The explanation of why this transition like
occurrence happens at about class O8 we offer at Fig.1. The figure
gives the average lifetime of stars in the mass range $1-100
\mathcal{M}_{\odot}$. For classes later that O8 the lifetime
$\tau$ starts to significantly increase, which gives them enough
time to abandon their birthplace.

\begin{table}[ph]
\tbl{Bright O stars and nebulosity from Ref. 6.
  }
  {\begin{tabular}{@{\extracolsep{0.5mm}}lccccccc@{}}
  \hline
   Location && \multicolumn{6}{c}{Spectral type} \\
            && O5 & O6 & O7 & O8 & O9 & O9.5 \\
   \hline
   In bright H II region   &\hspace{0.5cm} &4&8&8&14&3&2 \\
   In faint H II region    & &2&2&2& 6&6&3 \\
   Not in H II region      & &2&6&3&15&16&12 \\
 \hline
\end{tabular}}
\end{table}

\begin{figure}
\begin{minipage}{\textwidth}
  \centering
\includegraphics[bb=166 275 450 508,width=\textwidth,keepaspectratio]{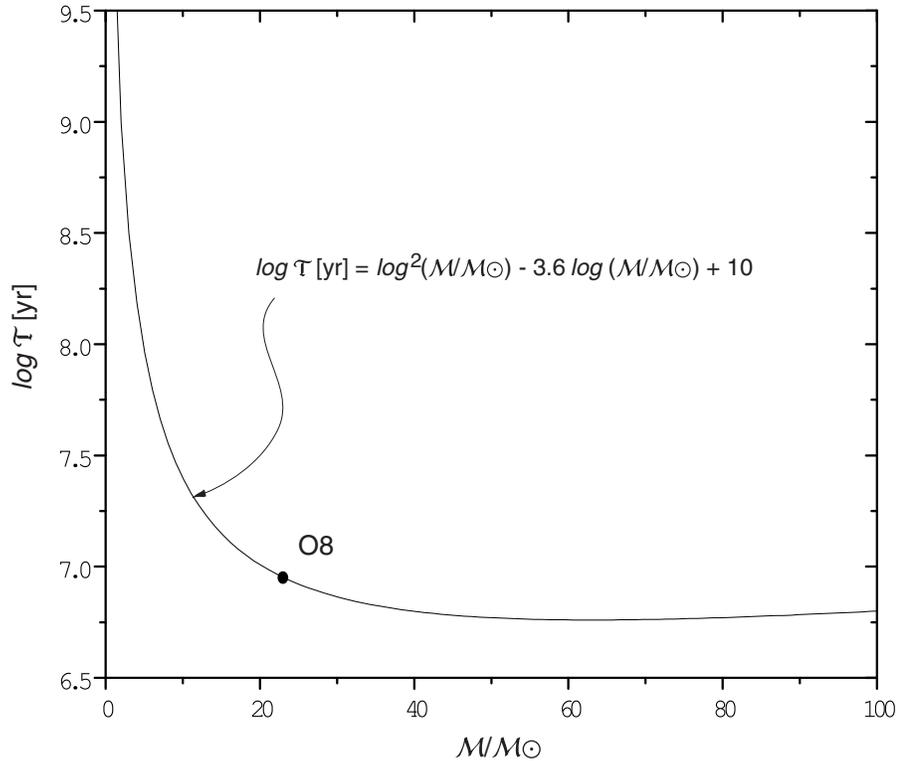}
  \caption{Stellar lifetime vs. initial mass in the range $1-100 \mathcal{M}_{\odot}$.
   For the relation adopted see Ref. 14.
   }
\end{minipage}
\end{figure}

The closest match to the obtained value for
$\bar{\mathcal{M}}_{\mathrm{Ib}}$ are exactly O8 stars with
$\mathcal{M} = 23 \mathcal{M}_{\odot}$ (from Allen's Astrophysical
Quantities, Ref. 15). Given the uncertainties in the SNRs we may
adopt the latter value as the lower limit for the
stripped-envelope core-collapse progenitors. The Ic progenitors
would then correspond to the stars earlier than O7 with masses
$\mathcal{M}_{\mathrm{Ic}}\gtrsim 30\mathcal{M}_{\odot}$.

Theoretical models for a single non-rotating star predict a much higher minimum SN Ib progenitor mass,
$\mathcal{M}_{\mathrm{Ib}}\gtrsim 30\mathcal{M}_{\odot}$.\cite{16}
Rotation seems to lower this value, $\mathcal{M}_{\mathrm{Ib}}\lesssim 25\mathcal{M}_{\odot}$\cite{17,18},
which is consistent with the empirical value obtained here.
On the other hand, the amount of mass loss suffered during the star's life, which will ultimately determine the type of supernova,
depends also on metallicity, and so does the minimum mass.\cite{17}
These three parameters: mass, metallicity and rotation, are the most important in single star supernova models.

A stronger association of SNe Ib/c and HII regions,
that might be expected if their progenitors are very massive stars, is, however,
questionable,\cite{19} and along with it the single star scenario.
Alternatively, stripped-envelope SNe may appear in close binary
systems (CBs) as a consequence of the Roche lobe overflow and the
process of mass transfer to the companion star.

According  to Ref. 20, presupernovae of SNe Ic thus may be the C-O
stars in CBs, formed after the two stages of mass transfer during
which H and He envelopes, respectively, had been lost. Similarly,
 SNe Ib may have progenitors
in the He stars in CBs (i.e. only the first stage of mass transfer
had been invoked).\cite{21}

Principally,  SNe Ib/c would then result from the same progenitor
population as most of the SNe II. Intuitively, however, Ic may
again be resulting from the more massive stars. Very provisionally
we may assume that, if the mass range for core-collapse SNe $8-100
\mathcal{M}_{\odot}$, is to be applied to the stripped-envelopes,
from
\begin{equation}
\frac{\int_{\mathcal{M}_L}^{\tilde{\mathcal{{M}}}_{\mathrm{Ib}}}
\mathcal{M} ^{-\beta}d\mathcal{M}
}{\int_{\tilde{\mathcal{{M}}}_{\mathrm{Ib}}}^{\mathcal{M}_U}
\mathcal{M}
^{-\beta}d\mathcal{M}}=\frac{\nu_{\mathrm{Ib}}}{\nu_{\mathrm{Ic}}}
\ ,
\end{equation}
$\tilde{\mathcal{{M}}}_{\mathrm{Ib}}$ would be around $\sim 10
\mathcal{M}_{\odot}$ (about class B2).

This is highly uncertain since there are many other important
parameters that must be taken into account for determining the
presupernova formation rate in CBs, such as the proximity of the
companion (and rotation and metalicity, important for a single
star also). Consequently there may be a considerable overlap in
the initial masses for these events. Perhaps these additional
parameters may also be related to the progenitor's mass, however,
 this demands a more detailed and rather complicated analysis
dealing with the formation and evolution of CBs.

What ever the exact supernova scenario is, only from the low Ib/c
SNR and the slope of the IMF, it may be conjectured that these are
quite rare events. We can simply follow the line of reasoning:
rare events -- unique physics -- tight constrains on progenitors.

\subsection{Supernova luminosities}

All supernovae, with the exception of some Ic (hypernovae), are
commonly believed to release about $\sim 10^{51}$ ergs in the form
of kinetic energy (excluding the neutrinos). In the case of the
hypernovae this number is believed to be larger, $\geq 10^{52}$
ergs. Only about 1 per cent of this energy is transformed into
light and radiated in a SN event.

Since this energy can be regarded as the released gravitational
potential energy of a star, it directly depends on the
presupernova's mass. SNe of a wide-mass-range progenitors are thus
likely to have a wider intrinsic luminosity distribution (such as
it is with SNe II). Tight constraints on progenitors (their mass,
metalicity, binarity, etc.) and a unique physics of explosion
means a smaller dispersion in observational properties. This may
apply to the luminosity of SNe Ib and SNe Ic. A sample of latter
would, however, still comprise the true hypernovae of
$>40\mathcal{M}_{\odot}$ stars.

 For the absolute magnitude at maximum (blue) light we
can generally write
\begin{equation}
\mathrm{M_B^0} = \mathrm{m_B} - \mu - \mathrm{A}_G - \mathrm{A}_g
= \mathrm{M_B} - \mathrm{A}_g
\end{equation}
where $\mathrm{m_B}$ is apparent magnitude, $\mu = 5\log
d\mathrm{[Mpc]} +25$ is distance modulus, $\mathrm{A}_G$ and
$\mathrm{A}_g$ are Galactic extinction and extinction in parent
galaxy, respectively. Ref. 22 gives the peak magnitude for Ib,
uncorrected for parent galaxy extinction $\mathrm{M_B}= -17.11 \pm
0.14$. This is consistent with the mean magnitude
\begin{equation}
\langle \mathrm{M_B} \rangle = -16.92 \pm 0.71 \approx -17,
\end{equation}
obtained from chosen sample of SNe Ib/c from the ASC (see Table
4). The mean error adopted in Ref. 22, however, seems too
optimistic. Since no extinction corrections were made, this value,
although it may be statistically useful, is surely fainter than
the true (intrinsic) magnitude.

The problem of extinction is the most important issue to be dealt
with, in the process of obtaining true SN luminosities (absolute
magnitudes).  The plane-parallel model which gives absorption
dependent on galaxy inclination $\mathrm{A}_g=\mathrm{A}_o
\mathrm{sec}\ i$, widely used in the past, was shown not to
describe extinction adequately.\cite{11} An alternative model
which introduces radial dependence was given in Ref. 23.

\begin{sidewaystable}
 \begin{minipage}{\textheight}
 {\small
  {{Table 4.} SNe Ib/c with listed B magnitude at the time of maximum from the
  Asiago Supernova Catalogue (ASC). The two SNe are excluded: SN 1954A because of the other unknown
  properties of SN, i.e. it's parent galaxy, and 1966J since it was shown to be
  Ia.\cite{24}
  All data are from ASC, except for distance moduli (Nearby
  Galaxy Catalogue - NGC, Ref.25) and correction for Galactic absorption $\mathrm{A}_G$ which is from RC3
   (Ref. 26).
  Parent galaxy extinction is omitted since we found significant
  discrepancy for $\mathrm{A}_o$ in RC3 and NGC. $\mathrm{M_B}$ is
  absolute magnitude uncorrected for extinction in the parent galaxy.
  }
  \vskip 5mm
  \centering
  \begin{tabular}{@{\extracolsep{-0.2mm}}lclcrrrrrrrrrrrrrrrrrrrrr@{}}
  \hline
   Supernova   & SN & Galaxy & Galaxy &
   \multicolumn{3}{c}{Distance}&
   \multicolumn{3}{c}{Inclination}&
   \multicolumn{3}{c}{Diameter}&
   \multicolumn{3}{c}{SN radial}&
   \multicolumn{3}{c}{Apparent}&
   \multicolumn{3}{c}{Galactic}&
   \multicolumn{3}{c}{Absolute}\\
           &  type &  & type &
   \multicolumn{3}{c}{modulus}&
   \multicolumn{3}{c}{}&
   \multicolumn{3}{c}{}&
   \multicolumn{3}{c}{position}&
   \multicolumn{3}{c}{magnitude}&
   \multicolumn{3}{c}{absorption}&
   \multicolumn{3}{c}{magnitude}\\
              & & & &
   \multicolumn{3}{c}{$\mu$}&
   \multicolumn{3}{c}{$i\ \mathrm{[^{\circ}]}$}&
   \multicolumn{3}{c}{$D\ \mathrm{[kpc]}$}&
    \multicolumn{3}{c}{$r\ \mathrm{[kpc]}$}&
   \multicolumn{3}{c}{$\mathrm{m_B}$}&
   \multicolumn{3}{c}{$\mathrm{A}_G$}&
   \multicolumn{3}{c}{$\mathrm{M_B}$}\\
\hline
SN 1972R & Ib & NGC 2841 & Sb && 30.39 &&\hspace{2mm} & 65 &&& 27 &&& 10.7 &&&  12.85  &&\hspace{1mm} &  0    &&& -17.54 &\\
SN 1983N & Ib & NGC 5236 & SBc&& 28.35 &&\hspace{2mm} & 21 &&& 18 &&& 3.8  &&&  11.70  &&\hspace{1mm} &  0.15 &&& -16.80 &\\
SN 1984I & Ib &ESO 393-99&SBcd&& 33.48 &&\hspace{2mm} & 25 &&& 30 &&& 11.0 &&&  16.60  &&\hspace{1mm} &  0.45 &&& -17.33 &\\
SN 1984L & Ib & NGC 991  & SBc&& 31.37 &&\hspace{2mm} & 28 &&& 16 &&& 3.5  &&&  14.00  &&\hspace{1mm} &  0    &&& -17.37 &\\
SN 2000H & Ib & IC 454   &SBab&& 33.89 &&\hspace{2mm} & 58 &&& 30 &&& 9.1  &&&  17.90  &&\hspace{1mm} &  1.44 &&& -17.43 &\\
SN 1962L & Ic & NGC 1073 & SBc&& 30.91 &&\hspace{2mm} & 25 &&& 21 &&& 5.7  &&&  13.94  &&\hspace{1mm} &  0.07 &&& -17.04 &\\
SN 1983I & Ic & NGC 4051 &SBbc&& 31.15 &&\hspace{2mm} & 35 &&& 26 &&& 5.5  &&&  13.70  &&\hspace{1mm} &  0    &&& -17.45 &\\
SN 1983V & Ic & NGC 1365 & SBb&& 31.14 &&\hspace{2mm} & 58 &&& 54 &&& 6.8  &&&  14.67  &&\hspace{1mm} &  0    &&& -16.47 &\\
SN 1987M & Ic & NGC 2715 & SBc&& 31.55 &&\hspace{2mm} & 74 &&& 28 &&& 2.1  &&&  15.30  &&\hspace{1mm} &  0.02 &&& -16.27 &\\
SN 1991N & Ic & NGC 3310 &SBbc&& 31.36 &&\hspace{2mm} & 19 &&& 15 &&& 0.8  &&&  15.50  &&\hspace{1mm} &  0    &&& -15.86 &\\
SN 1994I & Ic & NGC 5194 & Sbc && 28.40 &&\hspace{2mm} & 48 &&& 14 &&& 0.6  &&&  13.77  &&\hspace{1mm} &  0 &&& -14.63 &\\
SN 1998bw& Ic pec& ESO 184-82  & SB && 32.97 &&\hspace{2mm} & 33 &&& 9  &&& 2.6 &&&  14.09  &&\hspace{1mm} &  0 &&& -18.88 &\\
SN 2002ap& Ic pec& NGC 628     & Sc && 29.93 &&\hspace{2mm} & 24 &&& 30 &&& 13.7&&&  13.10  &&\hspace{1mm} &  0.13 &&& -16.93 &\\
 \hline

\end{tabular}
}
\end{minipage}
\end{sidewaystable}

\begin{figure}
  \begin{minipage}{\textwidth}
  \includegraphics[bb=166 276 449 508,width=120mm,keepaspectratio]{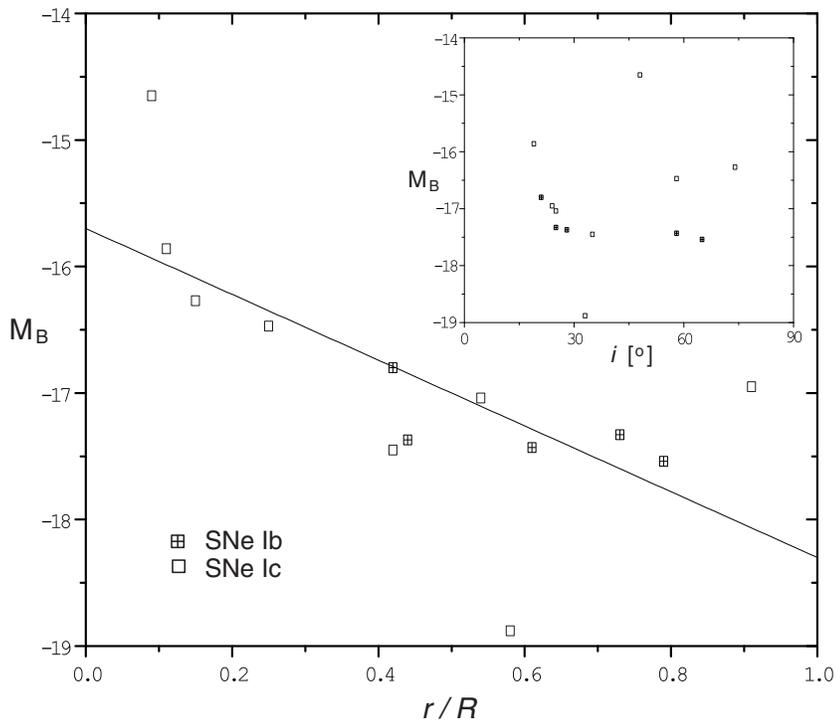}
  \caption
  {SN absolute magnitude, uncorrected for the parent galaxy extinction, is plotted against relative radial position of a SN in the galaxy. The straight line represents the least-squares fit.
  The inset in the upper-right corner of the plot shows, again, magnitude only now as the function of the galaxy inclination. There is no apparent dependence of $\mathrm{M_B}$ on $i$.
   }
  \end{minipage}
\end{figure}

Fig 2 shows peak magnitude uncorrected for parent galaxy
extinction against the radial position of SN $r$ in the units of
galactic radius $R=D/2$. The figure shows that there is a certain
trend of dimmer SNe with decreasing radius.
 If we assume that at $r/R=1$ extinction is negligible
intrinsic absolute magnitude for Ib/c SNe would be closer to
$\mathrm{M_B^0} = -18.31 \pm 0.45$.

It can also be seen on Fig 2 that SNe Ic do show larger
dispersion\footnote {One SN Ib with listed peak B magnitude in the
ASC, namely 1954A, was, however, not included because of the
unknown properties of SN, i.e. it's parent galaxy. The SN is in
the irregular galaxy NGC 4214. It is not near any HII region, and
it is untypically bright.}. The SNe that show the largest
deviations are 1994I, 1998bw and 2002ap. There are suggestions
that SN 2002ap, together with 1997ef and 1998bw, might even form a
new subclass Id (they may correspond to the hypernovae, connected
to the gamma-ray bursts).\cite{27} Probably the best studied SN
Ic, 1994 I, on the other hand may have incorrect distance caused
deviation.\cite{28}

\section{Conclusions}

The aim of this paper was to reanalyse the stripped-envelope
core-collapse SNe (Ib/c) and to try to bring them in connection to
their progenitors. If the isolated progenitor hypothesis is
correct, our analysis indicates that SNe Ib result from stars of
main-sequence mass $23 \mathcal{M}_{\odot} \lesssim
 \mathcal{M}_{\mathrm{Ib}} \lesssim 30 \mathcal{M}_{\odot}$, while the progenitors of SNe Ic are
more massive stars with $\mathcal{M}_{\mathrm{Ic}} \gtrsim 30
\mathcal{M}_{\odot}$.
 If majority of SN Ib/c, alternatively, appear in close binary systems
 then they would result from the same progenitor population as
 most of the
 SNe II, i.e. early B stars with initial masses of order
 $\mathcal{M} \sim 10 \mathcal{M}_{\odot}$.
 Analysis of this scenario, however, would have to
 include more details and will be rather complicated, dealing with the formation and evolution of CBs.

 The joint (Ib/c) intrinsic absolute magnitude
 obtained is
\begin{equation}
\mathrm{M_B^0} = -18.31 \pm 0.45 \approx -18,
\end{equation}
whereas SNe Ic are likely to show larger dispersion from this
value.

Future observations of SNe at high-redshift and their rate will
provide us with unique information on SN progenitors and
star-formation history of galaxies. At higher-$z$ (deeper in the
cosmic past) we expect to see the lack of type Ia events, i.e. the
dominance of core-collapse SNe. Better understanding of the
stripped-envelope SNe (Ib/c), and their potential use as distance
indicators at high-$z$, would therefore be of great practical
importance.\cite{29}

\section*{Acknowledgments}

The author would like to thank dr Dejan Uro\v{s}evi\'{c}, dr Milan
\'{C}irkovi\'{c} and prof. Sidney van den Bergh for reading and
commenting on the manuscript, and dr Enrico Cappellaro for
providing him with the list of SNe from his studies. During the
work on this paper the author was financially supported by the
Ministry of Science and Environment of Serbia through the projects
No. 146003 and 146012.

%\begin{thebibliography}{000} %for 3 digits
%\begin{thebibliography}{00}  %for 2 digits

\end{document}